# Room temperature observation of electron resonant tunneling through InAs/AlAs quantum dots

Jie Sun[a)], Ruoyuan Li, Chang Zhao, Like Yu, Xiaoling Ye, Bo Xu, Yonghai Chen and Zhanguo Wang[b)]

*Key Laboratory of Semiconductor Materials Science, Institute of Semiconductors, Chinese Academy of Sciences, P. O. Box 912, Beijing 100083, China*

Molecular beam epitaxy is employed to manufacture self-assembled InAs/AlAs quantum-dot resonant tunneling diodes. Resonant tunneling current is superimposed on the thermal current, and they make up the total electron transport in devices. Steps in current-voltage characteristics and peaks in capacitance-voltage characteristics are explained as electron resonant tunneling via quantum dots at 77K or 300K, and this is the first time that resonant tunneling is observed at room temperature in III-V quantum-dot materials. Hysteresis loops in the curves are attributed to hot electron injection/emission process of quantum dots, which indicates the concomitant charging/discharging effect.

Self-assembled quantum dots (QDs) grown in Stranski-Krastanow (S-K) mode have attracted much interest due to their potential applications in novel nanoscale devices[1]. However, with their optical properties having been well clarified, relatively less researches are done on transport through QDs, such as the fabrication of QD resonant tunneling diodes (RTDs). RTDs have the advantage over conventional circuits[2] in terms of reduced circuit complexity for implementing a given function. Since the quantum dots confine electrons in all three dimensions, resonant tunneling into the quantized states will give rise to additional peaks in current-voltage (I-V) characteristics[3] compared with quantum well RTDs. Furthermore, QD RTDs will operate with low power at high speed on account of their small sizes of devices. Therefore, QD RTDs should have a more promising future than quantum well RTDs. To date, however, all commercial RTDs are made of quantum wells whilst much more efforts are needed to make the quantum-dot RTDs practical. Room temperature observations of electron resonant tunneling via Si/SiO$_2$ QDs[4] or InGaN/GaN QDs[5] have been successfully achieved by virtue of the very large barrier height of matrix materials. However, a clear current bump due to resonant tunneling through III-V QDs e.g. InAs QDs can only be detected at extremely low temperatures[6-9], typically 1.6K-130K. Besides, nearly all this kind of experiment is focused on current-voltage characteristics of the samples, whereas little attention is paid to capacitance-voltage (C-V) characteristics. In the present letter, for the first time, clear step structures in the I-V curves of QD RTDs are observed at room temperature, and explained in terms of resonant tunneling via an ensemble of InAs/AlAs QDs. Also studied by C-V measurements here is the concomitant resonant charging effect. Hysteresis loops in I-V and C-V curves provide clues of hot electron charging and discharging procedure, thus further convincing us that as-grown samples show quantum-dot behavior. This finding represents a great step forward towards the applications of QD RTDs.

Our RTD samples, shown schematically in Fig. 1 (a), are grown by a RIBER 32P solid-source molecular beam epitaxy (MBE) machine on GaAs (100) substrate ($n^+=2\times10^{18}cm^{-3}$). The growth is started at substrate temperature $T_s$ of 600°C by depositing a 300nm n-GaAs, in which the Si concentration is graded from $2\times10^{18}cm^{-3}$ to $1\times10^{17}cm^{-3}$, and a 10nm undoped GaAs spacer layer. Then, a 5nm undoped AlAs barrier is grown and $T_s$ lowered to 530°C. At that point, nominally 1.9 monolayers (ML) InAs are deposited with a very slow rate of 0.015ML/s (the In delivery is cycled in 1.7s of evaporation followed by 5.0s of interruption until the given InAs thickness is reached) and a 2 min post-growth annealing[10] is processed. Low deposition rate, growth interruption and post-growth annealing all help to equilibrate the surface by enhancing the migration of In adatoms[11], consequently producing high quality QDs. After the QDs are capped by another 5nm undoped AlAs, $T_s$ is increased to 600°C. 10nm undoped GaAs spacer and 300nm graded n-GaAs ($1\times10^{17}cm^{-3}$ to $2\times10^{18}cm^{-3}$) are deposited in turn. With standard photolithography and wet etching procedures being used to obtain 60μm×60μm mesas, alloyed Au/Ge/Ni contacts are fabricated via small windows (40μm×40μm) in 350nm $SiO_2$ insulation layers. The sheet density of QDs is ~$6\times10^{10}cm^{-2}$ as determined from plan-view transmission electron microscope (TEM) measurements. It is known that upon AlAs capping at high growth temperature, In atoms will segregate towards the exterior to lower surface energy (the segregation ratio is 0.77 at 530°C, which is only slightly smaller than 0.85 in the InAs/GaAs system[12]). The upper AlAs barrier in our device, therefore, is in fact $In_xAl_{1-x}As$ ternary alloy. It is for this reason that the height of upper barrier has been lowered to some extent, as is illustrated schematically in Fig. 1 (b). Such effects will greatly influence the electronic properties of RTDs, which will be set forth thereinafter.

Summarized in Fig. 2-4 are room temperature and 77K I-V and C-V characteristics of three typical samples (A, B, C). It is with a crystal-controlled 1MHz 15mV test signal that the C-V properties are measured. C-V curves at 300K are shifted by 10pF for clarity. All curves are reproducible and the arrows indicate the voltage-scanning directions. All of the voltage sweeping begin and end at large negative bias. For all devices, the current at forward (positive) bias is smaller than at reverse (negative) bias. These asymmetric I-V characteristics reflect the intrinsic properties of the devices, which are attributed to inhomogeneity between Ohmic contacts of both sides and between the two asymmetric AlAs barriers [see Fig. 1 (b)]. Similar to previous studies[8, 9], the basic shapes of our I-V curves are determined by the hot current, and the resonant tunneling current which overlies on main structures only contributes to a small fraction of total current. Resonant tunneling signals (steps in I-V and peaks in C-V), at 77K or room temperature, are clearly observed in the figures and summarized in Table 1. Structures at 0.5—0.7V, -0.7—-1.1V are attributed to resonant tunneling through ground states of QDs, and signals at 2.1—2.8V, -1.9—-2.6V to first excited states. The cause for the broadening of I-V steps and C-V peaks is to be detected in the size undulation and lateral coupling of QDs[13]. It is well known that in resonant tunneling experiments, electrons will accumulate between double barriers[14], and that is why one can observe peaks in C-V properties when resonance takes place. This resonant tunneling charging effect will be enhanced in asymmetric double-barrier structures, whose reasons go as follows. In asymmetric double-barrier structures[15, 16], such as in our case [see Fig. 1 (b)], under forward bias the emitter barrier is less transparent than the collector one. That is, $T_E \ll T_C$, where $T_E$ and $T_C$ are electron transmission coefficients of emitter and collector barriers respectively. Therefore, there are relatively less electrons which remain in QDs while the majority passes through double barriers during resonant tunneling. When it comes to the reverse bias (in this case the emitter barrier is the upper $In_xAl_{1-x}As$ barrier), $T_E \gg T_C$, the electron density in the dots is considerably enlarged and resonant tunneling current drastically diminished. In other words, resonant tunneling signals in I-V and C-V will form a complementary pair. In Table 1, generally speaking, under negative bias the steps in I-V are non-salient whereas the corresponding C-V peaks strong enough to survive at 300K, indicating a strong resonant tunneling charging effect. When it comes to the positive bias, on the contrary, with the I-V stairs being obvious the C-V structures are hard to be found except at ~2.2V in sample B. It means that the resonance current is increased and the resonant tunneling charging effect decreases. Compared with sample A which does not exhibit any resonant tunneling signals at small positive bias (<2V) in I-V curve, one may notice how big the resonant tunneling current in samples B, C is. The step at 0.5V in I-V curve of sample C even persists up to room temperature.

As is mentioned above, the total current in our devices consists of two components: thermal current which determines the basic shape of I-V curve, and resonant tunneling current which generates steps superimposed on the main body. The latter having been discussed in detail, we will now focus on the former. I-V hysteresis loops, which are hints of hot electron charging effect (please note this is a completely different mechanism from the resonant tunneling charging effect investigated in the last paragraph) in QDs, are detected at liquid nitrogen temperature. Usually, when a sample with embedded QDs is biased, there may be two opposite consequences. One is charging of QDs via trapping hot electrons[17] (viz hot electrons relax into the QDs), and the other discharging of QDs by direct tunneling out through the barrier[18]. The dominant one will be selected through the competition between the two procedures. When I-V measurements start from big reverse bias (absolute value >1V), many QDs will capture relaxed hot electrons because thermal current is quite large. The electrons can not tunnel out easily since the collector barrier is high. These negatively charged QDs induce repulsive electrostatic potentials and act as scattering centers which conspicuously reduce the following current. The current will therefore be at its low states. In this stage, the C-V peaks also appear lower, for the resonant tunneling current lays aside less electrons into QDs due to the Coulomb repulsive effect from the hot electrons captured by QDs. When large positive voltages (>2V) are reached, the captured electrons are obliged to tunnel out[19] via the relatively lower collector barrier, and thereafter the current and capacitance alter to their high states in reverse scanning direction. The charging process at big negative voltage and discharging process at big positive voltage are schematically illustrated in Fig. 1 (b) by arrows (the energy band tilt under bias is ignored for simplification). The I-V hysteresis phenomena vanish at room temperature because of the limitedness of electron storage time. The electrons can easily and swiftly escape from QDs because of their large thermal energy at high temperature. Nevertheless, the loops do remain in C-V at 300K, which implies in that case the Coulomb repulsive forces come from hot electrons trapped *then and there* into the QDs. That conjecture can be further confirmed by the 300K C-V curve of sample B. When it scans from 0V to 3.3V, since there is almost no Coulomb repulsive forces (the large-negative-bias-trapped hot electrons have already escaped due to their transient lifetime in QDs at 300K), C-V resonant tunneling charging peak at 2.2V will be relatively high. While scanning from 3.3V back to 0V, the QDs may seize electrons from the much enhanced forward hot current at 300K. Therefore, the C-V peak at 2.2V will be lowered. In contrast to the 77K status where an anticlockwise hysteresis loop α is observed, the C-V loop β at 300K is clockwise. Therefore, it can be concluded that at room temperature the repellence does not arise from large-negative-bias-trapped[17] hot electrons, but from the real time trapped ones.

Finally, the discussion will go to several controversial issues. a) The detected signals are not because of resonant tunneling via the X energy band valleys in AlAs barriers. In similar researches, resonant tunneling through the X states of AlAs layers is usually not observed[8, 20, 21], indicating the X valleys of AlAs usually do not play a significant role in such kind of experiments. The resonance peaks in C-V characteristics and hysteresis loops in I-V and C-V curves all indicate a pronounced charging and discharging effect in our devices, and they strongly suggest that it is a quantum-dot behavior. b) More often than not, some abnormal phenomena, such as a swinging drop of capacitance before overflow from range, will be observed in C-V measurements when the conduction current is *too large*[22]. The mechanism is still arguable[23]. While the current passing through devices is too large, therefore, C-V measurements seem to be unreliable. That might explain why the hot electron injection/emission effect and the resonant tunneling peaks through QDs' first excited states (except at 2.2V in sample B) are not embodied in our C-V curves. c) In some previous researches[9, 24], not all the devices show resonant tunneling behaviors. Under our conditions, resonance signals can be found in all RTDs. However, there is still a certain kind of inhomogeneity among devices. The differences among sample A, B, C are not understood at present and more tests are needed to delve into the mystery.

We have, in short, fabricated InAs/AlAs QD RTDs for I-V and C-V studies. The total current passing through the devices is composed of hot electron current and resonance current. The observed stairs and peaks are attributed to electron resonant tunneling through QDs at 77K or 300K. To the best of our knowledge, this is the first direct observation of such phenomena at room temperature. Hysteresis loops in I-V or C-V curves are related to hot electron injection or emission via QDs, thus further convincing us that all the structures found in this experiment are quantum-dot behaviors. It is believed that this work will contribute to the potential applications of InAs QD RTDs.

Thanks go to Professors Hongqi Xu, Jianbai Xia, Yangyuan Wang and Jilin Gao for their generous support. Part of the work is finished with the help of Mrs. Chunli Yan. The financial aid is acknowledged highly valuable from Special Funds for Major State Basic Research Project of China (No 2000068303).

*Figure and table captions*

Fig.1(a). Schematic picture of the epitaxially grown RTD structure.

Fig.1(b). Depiction of band structure of the diode. The upper AlAs barrier is unintentionally changed into $In_xAl_{1-x}As$ alloy layer due to the In segregation effect during MBE growth. This makes the double barriers asymmetric. The upper arrow indicates the hot electron injection process at large negative voltage. The lower arrow indicates the hot electron tunneling out (i.e. emission) procedure at large positive voltage. Note the energy band tilt under bias is ignored in this figure for clarity.

Fig.2. Current-voltage and capacitance-voltage characteristics of sample A at 77K or 300K. The directions of voltage sweeping are shown by arrows. The scan begins and ends at -3V.

Fig.3. I-V and C-V curves of sample B. Hysteresis in the curves are caused by hot electron charging/discharging process. Note that loop α and β are of opposite directions.

Fig.4. I-V and C-V curves of sample C. The I-V curves are dominated by thermal current, with resonant tunneling current being superimposed on. The C-V peaks are due to so called resonant tunneling charging effect.

Table 1. Resonant tunneling signals in I-V and C-V properties of three typical samples at different biases. G and E stand for ground states and first excited states of QDs respectively. In general, I-V and C-V signals form a complementary pair.

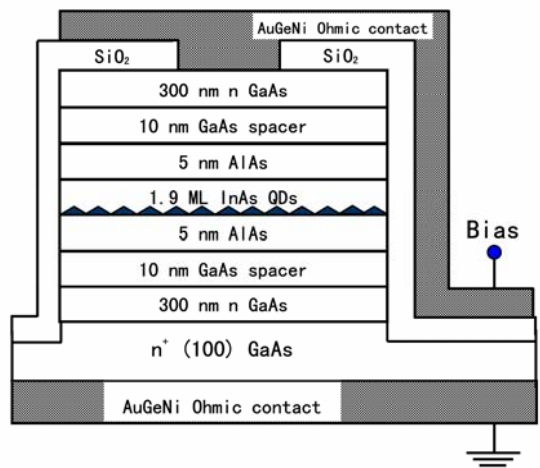

**Figure 1 (a)**

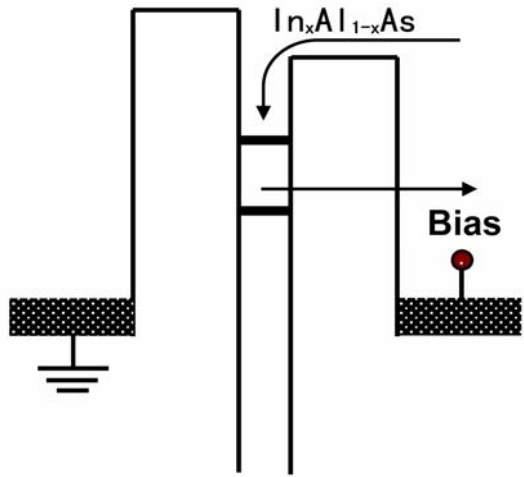

**Figure 1 (b)**

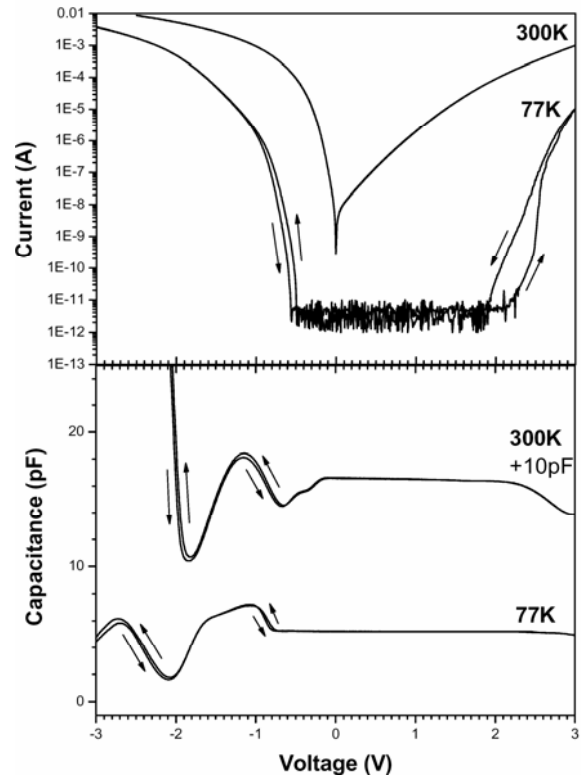

**Figure 2**

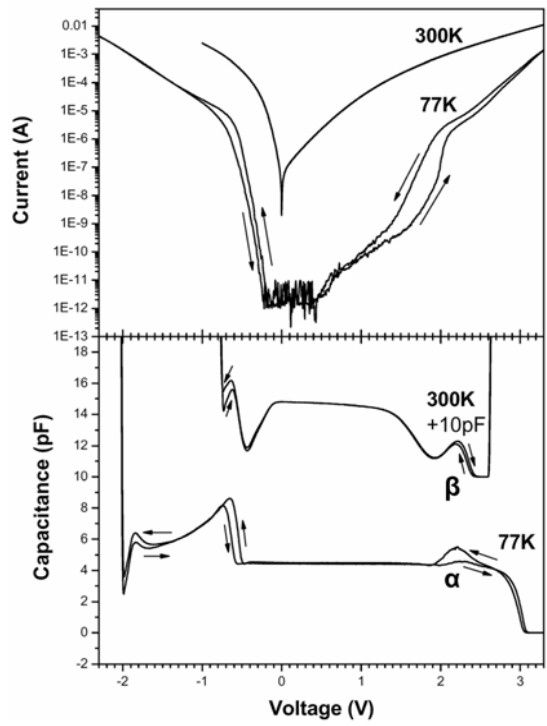

**Figure 3**

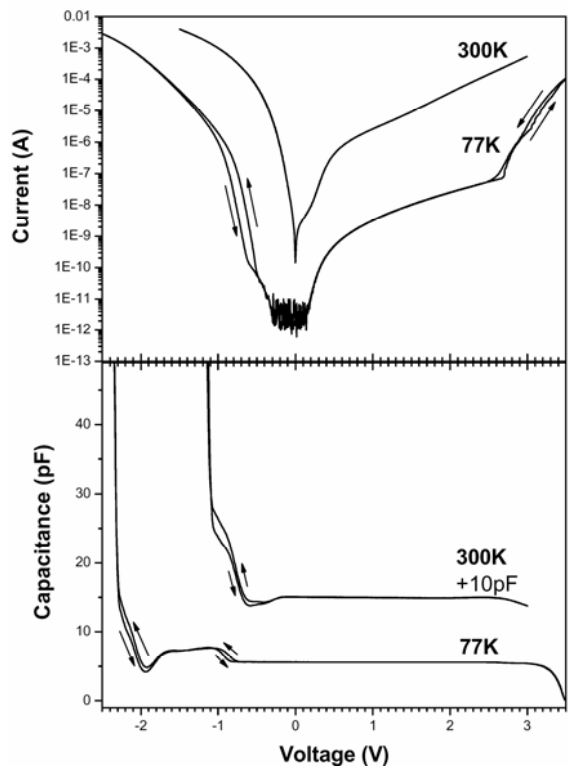

**Figure 4**

| Unit: V | Forward bias | | | | | | | | Reverse bias | | | | | | | |
|---|---|---|---|---|---|---|---|---|---|---|---|---|---|---|---|---|
| | I-V | | | | C-V | | | | I-V | | | | C-V | | | |
| | 77K | | 300K | | 77K | | 300K | | 77K | | 300K | | 77K | | 300K | |
| | G | E | G | E | G | E | G | E | G | E | G | E | G | E | G | E |
| Sample A | — | 2.5 | — | — | — | — | — | — | — | — | — | — | -1 | -2.6 | -1.1 | — |
| Sample B | 0.7 | 2.1 | — | — | — | 2.2 | — | 2.2 | -0.7 | — | — | — | -0.7 | -1.9 | -0.7 | — |
| Sample C | 0.5 | 2.8 | 0.5 | — | — | — | — | — | -0.8 | — | — | — | -1 | -2.1 | -0.9 | — |

**Table 1**